\documentclass[preprint,11pt]{elsarticle}

\usepackage[T1]{fontenc}
\usepackage[utf8]{inputenc}
\usepackage{url}
\usepackage{ae,aecompl}
\providecommand{\abs}[1]{\lvert#1\rvert} 
\providecommand{\norm}[1]{\lVert#1\rVert} 

\usepackage{graphicx}	
\usepackage{amsmath}	
\usepackage{amssymb}	
\newtheorem{theorem}{Theorem}[section]
\newtheorem{definition}{Definition}[section]

\usepackage{color}

\journal{Physica D}

\begin{document}

\begin{frontmatter}



\title{Ill-posedness of the mean-field dynamo equations with a linear electromotive force}

\author[inst1]{M. E. Rubio}

\affiliation[inst1]{organization={Instituto de Astronomía Teórica y Experimental (IATE-CONICET)},
            addressline={Laprida 854}, 
            city={Córdoba},
            country={Argentina}}

\author[inst1,inst2]{F. A. Stasyszyn}

\affiliation[inst2]{organization={Observatorio Astronómico de Córdoba (OAC-UNC)},
            addressline={Laprida 854}, 
            city={Córdoba},
            country={Argentina}}

\begin{abstract}
We show that the initial-value problem for the non-relativistic magnetic dynamo equation turns out to be ill-posed in $L^2$ norm when the electromotive force depends linearly on the magnetic field. This result implies that the increasing of magnetic energy does not necessarily come from physical amplification mechanisms, since certain magnetic modes could arbitrarily grow as wave-frequency increases, despite any dynamo process. Thus, up to this order, the theory is not suitable for astrophysical simulations. We then study the case when electromotive forces are linear in magnetic field derivatives, showing that the resulting system has a well-posed problem. Finally, we apply the well-posed theory to the force-free regime, for which we find bounds for the corresponding magnetic energy analyzing the evolution of the magnetic helicity.
\end{abstract}


\end{frontmatter}



\section{Introduction and motivations}

One of the most challenging open problems in modern Astrophysics is, undoubtedly, to determine the origin of magnetic fields in galactic and extragalactic scales \cite{Brandenburg18,Kunze13,Brand20Prim}. The study and detection of magnetic fields in galaxy clusters has attracted much attention during the last ten years, showing a significant progress in their detection on galactic halos \cite{Krause14,Beck19,Kahniashvili16,Kahniashvili18}, although successful measurements in larger scales, such as filamentary structures, are still missing \cite{Colgate00,Ryu98}. 

Much effort has been devoted into a better understanding of the evolution and organization of magnetic field lines over larger scales. Theoretical and numerical tools have been developed, allowing a huge variety of highly accurate MHD simulations \cite{Miesch15,Roper20,Schober18,Brandenburg19SolarEvo,Kapyla20}. Some of them suggest that magnetic field saturates after reaching the corresponding equipartition value in the halos of astrophysical objects, being its intensity dependent on the seed field \cite{Brandenburg18}. This idea is in tension with other hypothesis claiming that saturation of the interstellar magnetic field is actually secondary to its origin \cite{Kulsrud07}, opening a wide range of speculations about the mechanisms from which magnetic fields get amplified. Some approaches involve a variety of statistical methods for a deep analysis of rotation measurements in large-scale structures, based on data that is expected to be obtained with the new generation of radio-telescopes \cite{Taylor12,Schilizzi11}. Nevertheless, there is no concrete evidence of the presence of magnetic fields on the surface of last scattering \cite{Widrow02}, giving rise to the fundamental question on when did the first magnetic fields arise. This still remains unanswered, and motivates part of this work. 

It is common in the literature the hypothesis that the maintenance and amplification of large-scale magnetic fields are achieved by \textit{dynamo}-type mechanisms \cite{Kronberg94,Widrow02,Widrow12}, by which magnetic field is continuously regenerated by differential rotation and helical turbulence. This is not the case, nevertheless, for slowly rotating systems (such as galaxy clusters), in which the fields have a characteristic scale much smaller than the whole size of the system, making thus their organization in larger scales a more difficult process.

Roughly speaking, a \textit{magnetic dynamo} consists of electrically conductive matter that moves in an external magnetic field, such that the induced currents can maintain and even amplify the original field. A few decades after Larmor's suggestion about dynamo processes as responsible for astrophysical magnetic fields, Steenbeck, Krause and Radler focused on the importance of \textit{helical turbulence} for dynamos in stars and planets \cite{Pouquet76}. These ideas were soon applied to the problem of galactic magnetic fields \cite{Dormy07,Parker70,Vainshtein71} in which a standard galactic dynamo model known as $\alpha\omega$-dynamo emerged (see \cite{Widrow02} for a nice review).
Although dynamo-type mechanisms are widely accepted as primary for the maintenance of magnetic fields in celestial bodies such as the Sun and also in galaxies, such a hypothesis for extragalactic scales is a bit more speculative. However, it may be plausible that dynamo processes operate ``hierarchically'' from sub-galactic to galactic scales, given its ability to continuously regenerate large-scale magnetic fields. Observational methods mostly focus on synchrotron emission, Faraday rotation, Zeeman splitting, and polarization of optical starlight \cite{Beck19}, getting measurements within the intracluster medium of about a few $\mu G$ \cite{Widrow02,Widrow12}, thus reaching almost the same values for typical galaxies \cite{Fletcher2011}. Likewise, the ``dynamo paradigm'' as a mechanism for maintaining, amplifying or regenerating magnetic fields should be considered incomplete for several reasons. As an example, the temporal scale for the amplification of the fields could be too long in order to explain their observation in younger galaxies, not necessarily revealing the origin of initial fields as seeds for subsequent dynamo action.

The high complexity of the equations governing the time evolution of magnetised plasmas only allow for exact solutions under rather simplified assumptions, needing to resort to computer-based simulations for obtaining reliable results. Numerical models for galaxy formation and evolution are extremely demanding from a computational point of view, even when neglecting magnetic fields, requiring high-performance computing, spanning over a wide spatial range, for instance from a parsec up to million of them. Advances in technology allowed a continuous increase in the computing performance, particularly in terms of parallelisation, becoming more viable to take into account all demands imposed by the MHD equations.

From a numerical perspective towards simulating astrophysical dynamos, a rigorous inspection of the initial-value problem for the corresponding system of equations turns out to be essential. As we shall prove in this work, the currently accepted dynamo theory might admit models on which the increasing of magnetic energy could not necessarily come from physical amplification mechanisms. It is not difficult to consider systems described by dynamical equations that actually amplify the magnetic field in arbitrarily large orders of magnitude, despite any astrophysical process. This ``anomaly'' is purely related to mathematical properties of the evolution equations, which may admit unphysical modes due to the non-diagonalizability of its principal part. This behavior may cause, thus, an arbitrarily fast increasing of magnetic energy, being able to be present in numerical simulations. When something like this happens for an evolution system of equations, we refer the system as to be \textit{ill-posed}, since it is not possible to find any norm with which to control the evolution with respect to the initial data, being rather impossible to predict any further dynamics (not even guarantee uniqueness of the solution). The notion of well-posedness results intrinsic and substantial for the description of a physical system, since it helps to choose better theories avoiding these type of anomalies.

In this work we address a detailed analysis of some mathematical and physical properties of the system of equations that model the evolution of magnetic fields under the \textit{mean field approximation} \cite{Brandenburg18}. In particular, our study concerns the \textit{hyperbolicity} of the magnetic dynamo equation: a crucial tool for guaranteeing a well-posed initial-value formulation of the theory. As we shall review later on, hyperbolicity implies uniqueness of the solution given certain initial data set, as well as continuous dependence of the evolution with respect to them. The relevance of inspecting the hyperbolicity of these equations takes center stage when carrying out fully numerical simulations on astrophysical dynamos, since what is sought is to avoid the propagation and arbitrary growth of certain undesired perturbations that do not represent dynamo-like processes; rather, they arise as product of setting an unsuitable system of dynamical equations.

\subsection{Outline and conventions}

This work is organized as follows. In section \ref{sec-2} we revisit the magnetic induction equations and introduce the dynamo equation that shall be further studied. Section \ref{sec-3} is devoted to the analysis of the hyperbolicity of the magnetic dynamo equation with two different choices for the electromotive force. The simplest one considers only a linear magnetic field dependence, showing that the corresponding initial-value problem is ill-posed, in contrast to the second choice, which includes linear contributions from magnetic field derivatives. In section \ref{sec-4}, these results are used in the context of force-free dynamos. After showing that the corresponding constraint equations properly propagate, we derive estimates for the magnetic energy, using an identity that is also shown. Comments and concluding remarks are contained in section \ref{sec-5}. Finally, a brief discussion concerning well-posedness of linear systems in the context of the problem here addressed is given in \ref{sec-ap}. 

Throughout all this work we shall consider geometric units such that $c = G = 1$, where $c$ is the speed of light in vacuum and $G$ is Newton's gravity constant.

\section{The magnetic dynamo equation}	\label{sec-2}

The induction equation is a powerful tool to model the dynamics of a wide variety of physical phenomena involving magnetized plasmas at different scales. In ideal magnetohydrodynamics (MHD), this equation reads \cite{Moffatt78}
\begin{equation} \label{ideal-mhd}
\partial_t \vec{B} = \vec{\nabla}\times(\vec{v}\times\vec{B}),   
\end{equation}
where $\vec{B}$ is the magnetic field, and $\vec{v}$ is the fluid velocity. Equation (\ref{ideal-mhd}) is a straightforward consequence of combining Faraday’s law
\begin{equation}
    \partial_t \vec{B} = - \vec{\nabla}\times\vec{E}
\end{equation}
and Ohm’s law for an ideal conductor, namely
\begin{equation}\label{ideal-ohm}
\vec{E} = - \vec{v} \times \vec{B}.
\end{equation}
For finite-conductivity systems (i.e., \textit{non-ideal} conductors), equation (\ref{ideal-ohm}) generalizes to
\begin{equation}
    \vec{E} = - \vec{v} \times \vec{B} + \frac{1}{\mu_o\sigma}\vec{\nabla}\times\vec{B},
\end{equation}
where $\sigma$ is the conductivity, and equation (\ref{ideal-mhd}) becomes
\begin{equation}\label{dif-induction-eq}
  \partial_t \vec{B} = \vec{\nabla}\times\left(\vec{v}\times\vec{B} - \eta \vec{\nabla}\times\vec{B}\right),  
\end{equation}
commonly known as the \textit{diffusive induction equation}, with 
\[
\eta = \frac{1}{\mu_o \sigma}
\]
is the \textit{magnetic diffusivity}.

Equation (\ref{dif-induction-eq}) has been deeply studied in the past. Theoretical and numerical tools have been developed towards a better understanding of the evolution and organization of magnetic field lines at different regimes, even allowing a huge variety of highly accurate MHD simulations (see \cite{Brandenburg18} for a nice a complete review). This requires, undoubtedly, a well-posed initial-value formulation. 

Moreover, one is often needed to resort to a generalization of equation (\ref{dif-induction-eq}), known as the \textit{magnetic dynamo equation} for the \textit{mean} magnetic field, namely
\begin{equation}\label{dynamo-eq}
  \partial_t \vec{B} = \vec{\nabla}\times\left(\vec{v}\times\vec{B}\right) + \vec{\nabla}\times\vec{\mathcal{E}}(B,\partial B,\partial^2 B,\cdots).
\end{equation}
This equation is driven by an \textit{electromotive force}, $\vec{\mathcal{E}}$, which may depend linearly on $\vec{B}$ as well as on their spatial derivatives \cite{Brandenburg18,Kunze13,Beck19}. Different choices for the electromotive force give rise to different dynamical properties the equations satisfy along evolution. The simplest choice for studying dynamos is when $\vec{\mathcal{E}}$ is a linear function of $\vec{B}$ via the \textit{mean helicity} field. Surprisingly however, an interesting evidence on the failure to consider the magnetic dynamo equation with an electromotive force which is purely linear with the magnetic field has been pointed out in \cite{Yoshizawa90,Yokoi13}. Although a wide range of formulations/proposals came up during the last decade on how to model this term, some subtle aspects regarding the hyperbolicity of equation (\ref{dynamo-eq}) have not been pointed out before, to the best of our knowledge, which motivates this work. 

Here we address the initial-value problem of equation (\ref{dynamo-eq}), with the two simplest choices for the electromotive force $\vec{\mathcal{E}}$. One of the motivations of our study lies in finding suitable models for describing magnetic field amplification in cosmological filaments, and in particular, exploring velocity field profiles capable of amplifying magnetic fields from smaller to larger scales. This is one of the most challenging open problems in Modern Astrophysics, given the difficulty of the system of equations governing their dynamics. The aim of this note is, thus, to contribute to discarding some of these simplest choices for the electromotive force since, although they may give promising numerical results, are not physically reasonable, as we shall justify throughout this work.

The search of fluid fields that could allow an increase of the magnetic energy is in general a highly non-trivial task. The reason is that, in the most general picture, one is devoted on looking for solutions of the magnetic induction equation coupled with the dynamics of a fluid system, which can be modeled as satisfying Euler equations (in the simplest case), or the Navier-Stokes equations (even any other dissipative fluid theory) if one is interested in including energy transport mechanisms. Nevertheless, in order to prove that a full system of equations is \textit{ill-posed}, it is enough to consider just a rendition of it, or what is generally known as the \textit{kinematic} regime \cite{kreiss2004initial, Strang66, Reula04}. This means that, if the system of equations is \textit{ill-posed} in the \textit{kinematic} regime, then it will be so in the full general case. Particularly, in this work we shall concentrate our analysis in the evolution equations for the magnetic field, assuming a background fluid solution (usually a stationary solution) and show that rendition to constitute an ill-posed initial-value problem. This result will directly imply that the full ``\textit{dynamo $+$ fluid}'' system will share the same mathematical property.

\section{Hyperbolicity}
\label{sec-3}

\subsection{Preliminaries}

Modeling physical phenomena through theories helping to predict their subsequent dynamics leads to looking after a systematic treatment of the dynamical fields and the set of equations they satisfy.
Surprisingly, there appear common patterns  which are closely related to the mathematical structure over which the theory is defined. Dynamical evolution is determined by certain set of fields $\left\{\varphi^{\alpha}\right\}$ defined over certain spacetime $(\mathcal{M},g)$, satisfying some system of equations $\mathcal{G}[\varphi^{\alpha}] = 0$, together with what is known as the \textit{initial-value formulation}. Generally, initial data cannot be given arbitrarily, since they must satisfy certain set of \textit{constraint equations}; i.e., differential equations in which only spatial derivatives appear, which must be satisfied at each time during further evolution. The initial-value problem is defined, thus, by prescribing the value of the fields on some spatial hypersurface $\Sigma_o$ \cite{friedrichs1971systems,geroch1996partial}.

There are three conditions that any theory must satisfy in order to admit a \textit{well-posed} initial-value formulation \cite{Hadamard1908}: (i) \textit{existence} of a solution; (ii) \textit{uniqueness} of such a solution, and (iii) \textit{continuous dependence} with respect to the initial data. Condition (i) is clear; condition (ii)--although often essential to establish mathematical properties about the solution-- is related to other two fundamental aspects: the \textit{predictability power} of the theory (which clearly seeks to describe ``realistic'' phenomena) and the so-called \textit{causality principle}, which states that every plausible theory describing evolutionary processes should be consistent with the \textit{causal structure} of the spacetime on which it is defined. The corresponding dynamic evolution is governed by the \textit{principal part} of the equations, which contains information about the propagation speeds of the different modes \cite{Kreiss70}.

As it was motivated in the introduction, one of the fundamental concepts that arise when studying the evolution of dynamical systems is their \textit{hyperbolicity}, encompassing aspects of the theory that must be fulfilled even in the most fundamental scenarios \cite{Friedrichs54,Kreiss70,friedrichs1971systems,geroch1996partial}. In what follows, we particularize some of these ideas to the magnetic dynamo system of equations, and study the corresponding initial-value problem for particular choices of the electromotive force. In particular, given that the definition of well-posedness involves the existence of a norm in the function space of the solutions (see \ref{sec-ap}), our results throughout this work are given with respect to the $L^2$ norm, since it is the natural norm defined in the space of functions in which the solutions of interest belong to.

\subsection{The equations}

As it is well known, Magnetohydrodynamics is governed by Maxwell's equations (in appropriate limits), coupled with Hydrodynamics. In the most general case, hydrodynamic equations could take into account dissipative effects, energy and heat transport phenomena, and ``magnetic pressure'' terms. Nevertheless, the study of astrophysical dynamos make use of a mean-field approximation to describe the effects of turbulence, sometimes ignoring the backreaction of the magnetic field on the fluid, reducing the problem, thus, to a purely kinematic one \cite{Widrow02}. As pointed out before, we shall consider the dynamics of magnetic field due to the induction system of equations, assuming a given background flow.

In the mean field approximation \cite{Widrow12,Brandenburg18}, it is assumed that both the velocity and magnetic field are decomposed into a mean part ($\langle\vec{v}\rangle$ and $\langle\vec{B}\rangle$) which slowly varies on the characteristic large scale, say $L$, and a fluctuating part ($\vec{v}'$ and $\vec{B}'$) which rapidly varies in a smaller scale and such that $\langle\vec{B}'\rangle = \langle\vec{v}'\rangle = 0$. Considering a homogeneous,  isotropic,  and  non mirrorsymmetric  turbulence,  the set of dynamical equations for the mean magnetic field reads
\begin{equation}\label{induction}
\left\{
\begin{array}{rcl}
    \displaystyle \partial_t\vec{B} &=& \vec{\nabla}\times (\vec{v}\times\vec{B}) + \vec{\nabla}\times\vec{\mathcal{E}}(B,\partial B,\partial^2 B, \cdots)\\
    \\[-0.3cm]
    \vec{\nabla}\cdot\vec{B} &=& \mbox{0}
\end{array}
\right.
\end{equation}
where $\vec{B}(t,\vec{x})$ is the mean magnetic field, $\vec{v}(t,\vec{x})$ the corresponding background fluid and $\vec{\mathcal{E}}$ the electromotive force due to turbulent motions of the magnetic field as it is carried around by the fluid. In general, the electromotive force can be expressed as an expansion of terms which depend on spatial derivatives of $\vec{B}$ of arbitrary order, namely \cite{Widrow02}
\begin{equation}
    \mathcal{E}^i = \alpha^{ij} B_j + \beta^{ijk}\partial_j B_k + \gamma^{ijk\ell}\partial_j\partial_kB_{\ell} + \cdots
\end{equation}
where each election for tensors $\alpha^{ij}$, $\beta^{ijk}$, $\gamma^{ijkl}$, $\cdots$ will clearly lead to a different dynamic for the magnetic field. It is worthwhile to mention that the transport coefficients contained in each of those tensors can be explicitly evaluated for specific astrophysical plasma systems and even for the interplanetary space (see for instance \cite{Bourdin18,Narita18}).

In the mean field regime, and as first step towards a correct description of magnetic fields, we shall study the case in which the electromotive force is purely linear in $\vec{B}$, that is $\vec{\mathcal{E}} = \alpha \vec{B}$, where $\alpha$ is the \textit{mean helicity} of the background flow
\begin{equation}
    \alpha = - \frac{\tau}{3}\langle \vec{v}\cdot(\vec{\nabla}\times\vec{v})\rangle,
\end{equation}
$\tau$ is the correlation turbulence time, and $\langle \cdots \rangle$ denotes ensemble average. This corresponds to taking $\alpha^{ij} = \alpha \delta^{ij}$. After that, we consider the ``difussive'' case, namely
\begin{equation}
    \vec{\mathcal{E}} = \alpha \vec{B} - \beta \vec{\nabla}\times \vec{B},
\end{equation}
which corresponds to setting $\beta^{ijk} = -\beta \varepsilon^{ijk}$, where $\varepsilon^{ijk}$ is the Levi-Civita symbol in three spatial dimensions. The coefficient $\beta$ takes into account both molecular and turbulent magnetic difussivities \cite{Kulsrud07}, usually set to
\begin{equation}
    \beta = \frac{\tau}{2}\langle v^2\rangle.
\end{equation}

One of the indicators of magnetic field growth during evolution is the global \textit{magnetic energy}
\begin{equation}
    E_{M} = \frac{1}{8\pi}\int_{\mathbb{R}^3}{B^2},
\end{equation}
which is proportional to the square of the $L^2$ norm of the magnetic field.
Nevertheless, this quantity is not enough to compute the growth of magnetic field energy through magnetic field modes, as we shall point out in what follows.

\subsection{Well-posedness}

We now address the initial-value problem of system (\ref{induction}), in the cases in which the electromotive force is (i) linear in the magnetic field and (ii) linear in first-derivatives of the magnetic field. To do so, we analyze the principal part of the system in both cases, and study the existence of unphysical modes in the high-frequency limit that render the system non-hyperbolic (and thus, ill-posed).

\subsubsection{Ill-posedness for $\alpha\neq 0$, $\beta = 0$}
We start by analyzing the hyperbolicity of the equation
\begin{equation}\label{ind-eq-eta0}
    \partial_t \vec{B} = \vec{\nabla}\times (\vec{v} \times \vec{B}) + \vec{\nabla} \times (\alpha\vec{B}).
\end{equation}
Since we are coupling equation (\ref{ind-eq-eta0}) with the differential constraint $\vec{\nabla}\cdot\vec{B} = 0$, we need to check that it properly propagates along evolution. This is rather simple in this case, as defining $C_1:=\vec{\nabla}\cdot\vec{B}$ we get
\begin{equation}
    \partial_t C_1 = \vec{\nabla}\cdot \left[\vec{\nabla}\times (\vec{v} \times \vec{B} + \alpha\vec{B})\right] = 0,
\end{equation}
since div(rot($\cdot$)) = 0. Thus, if we choose $\vec{B}$ such that $C_1 = 0$ at $t=0$, then $C_1 \equiv 0$ for any further time. 

The principal part of equation (\ref{ind-eq-eta0}) is (see \ref{sec-ap} for definitions and conventions)
\begin{equation}
    \partial_t \vec{B} = \vec{\nabla}\times (\vec{v} \times \vec{B}) + \alpha \vec{\nabla} \times \vec{B}.
\end{equation}
We now look for wave-like solutions of the form
\begin{equation} \label{Bmodes}
\vec{B} = \vec{B}_o e^{i(\omega t + \vec{k}\cdot\vec{x})},
\end{equation}
from which we have $\partial_t \vec{B} = i\omega\vec{B}$, and the subsidiary equation reads
\begin{eqnarray*}
 \omega\vec{B} &=& \vec{k}\times(\vec{v}\times\vec{B}) + \alpha \vec{k}\times \vec{B} \nonumber \\
 &=& (\vec{k}\cdot \vec{B}) \vec{v} - (\vec{k}\cdot \vec{v}) \vec{B} + \alpha \vec{k}\times\vec{B}.
\end{eqnarray*}
Without loss of generality, we can choose a frame such that $\vec{k} = (k,0,0)$, where $k := |\vec{k}|$.
Thus, the subsidiary system of equations for the modes reads
\begin{eqnarray*}
\omega B_1 &=& 0 \nonumber \\
k v_2 B_1 - \left(k v_1 + \omega\right)B_2 - k\alpha B_3 &=& 0 \nonumber \\
k v_3 B_1 + k\alpha B_2 - \left(k v_1 + \omega\right) B_3 &=& 0 \nonumber
\end{eqnarray*}
or $\mathcal{M}\vec{B} = 0$, with
\begin{equation}
       \mathcal{M} =
  \left( {\begin{array}{ccc}
   \omega & 0 & 0 \\
   k v_2 & -\left(k v_1 + \omega\right) & - k\alpha \\
   k v_3 & k\alpha & - \left(k v_1 + \omega\right) \\
  \end{array} } \right)
\end{equation}
Since we are looking for nontrivial solutions, we ask for the algebraic condition
\begin{equation}
    \det\left(\mathcal{M}\right) = 0,
\end{equation}
which leads to the following dispersion relation:
\begin{equation}
    \omega \left[\left(k v_1 + \omega\right)^2 + (k \alpha)^2\right] = 0,
\end{equation}
with solutions
\begin{equation}
    \omega_o = 0, \qquad \omega_{\pm} = -k v_1 \pm ik\abs{\alpha}.
\end{equation}
Thus, there is a channel $\omega_- = -kv_1 - ik\abs{\alpha}$ such that the mode $\vec{B}_- \sim e^{-ikv_1t} e^{\abs{\alpha}k t}$ grows without bound in the high-frequency limit, and the principal symbol turns out to be not diagonalizable with purely real eigenvalues. 

Finally, from the above argument, we can also conclude that the full linear system of equations (with variable coefficients) is ill-posed, using a result provided by Strang in \cite{Strang66}. The author deals with more general linear systems, namely
\begin{equation} \label{eq:strang-syst}
\partial_t u = \sum_{\abs{\alpha}\leq m}{A_{\alpha}(x)D^{\alpha}u},
\end{equation}
where $x=(x_1,\cdots,x_n)\in\mathbb{R}^n$, $u=u(t,x)\in\mathbb{C}^s$, $\alpha := (\alpha_1,\cdots,\alpha_n)\in\mathbb{N}_o^{n}$ and
\[
D^{\alpha} := \frac{\partial^{\abs{\alpha}}}{\partial x_1^{\alpha_1}\cdots \partial x_n^{\alpha_n}}, \qquad \abs{\alpha} := \alpha_1 + \cdots + \alpha_n;
\]
asserting that, if system (\ref{eq:strang-syst}) is well-posed, then the system of equations that results by freezing-out the coefficients at some $x_o$ \textit{is also} well-posed. For general (quasi-) linear systems, the principal part coincides with the linearization at an arbitrary solution. Thus, it is enough to take a linearization of the full system around an equilibrium solution and prove that such rendition is not well-posed in order to guarantee the general ill-posedness of the full system, concluding the proof.

This implies that equation (\ref{ind-eq-eta0}) is non-hyperbolic and does not lead to a well-posed initial-value formulation. Moreover, this result holds despite the addition of the evolution equations for the fluid flow, being enough to show that this reduction is ill-posed to state the ill-posedness of the corresponding equations coupled to any fluid theory. This subtlety follows from a series of well-known theorems involving microlocal analysis and PDE theory (for further details on this aspect, we refer the reader to the work \cite{Reula04}, as well ass to \cite{Taylor91,Taylor12,Strang66}, in which a generalization of the notions of hyperbolicity for quasi-linear systems in a fully covariant way is developed). This result is quite relevant since it shows that it is not suitable to model dynamo processes just by taking an electromotive force which is linear in the magnetic field.

\subsubsection{Well-posedness for $\alpha\neq 0$, $\beta\neq 0$}

We now consider the full induction equation, up to quadratic magnetic field contribution for the electromotive force, namely
\begin{equation}\label{ind-eq-full}
    \partial_t\, \vec{B} = \vec{\nabla}\times (\vec{v} \times \vec{B}) + \vec{\nabla} \times (\alpha\vec{B}) + \beta\nabla^2\vec{B}.
\end{equation}
In this case, the constraint $C_1= \vec{\nabla}\cdot\vec{B}$ also propagates correctly, leading to the equation
\begin{equation}
    \partial_t C_1 = \beta\, \nabla^2 C_1,
\end{equation}
which is parabolic. Since $\beta>0$, by the uniqueness of this equation and setting the initial data such that $C_1(t=0) = 0$, we directly get $C_1 \equiv 0$ for any further time.

Following a similar analysis that the one performed in the previous case, we look for solutions of the form (\ref{Bmodes}). In this case, we arrive to the equation
\begin{equation}
    \left(\omega + \vec{v}\cdot \vec{k} - i\beta \abs{k}^2\right)\vec{B} = \alpha \vec{k}\times\vec{B}.
\end{equation}
We find it useful to introduce the function
\begin{equation}
    \Omega = \omega + \vec{v}\cdot \vec{k} - i\beta \abs{k}^2,
\end{equation}
from which the system now reads 
\begin{equation}
    \mathcal{N}\vec{B} = 0,
\end{equation}
where
\begin{equation}
       \mathcal{N} =
  \left( {\begin{array}{ccc}
   \Omega & \alpha k_3 & - \alpha k_2 \\
   -\alpha k_3 & \Omega & \alpha k_1 \\
   \alpha k_2 & - \alpha k_1 & \Omega \\
  \end{array} } \right).
\end{equation}
For the dispersion relation, we get
\begin{eqnarray*}
    0 &=& \det\left(\mathcal{N}\right) \nonumber \\
      &=& \Omega\left(\Omega^2 + \alpha^2 \abs{k}^2\right),
\end{eqnarray*}
with solutions
\begin{equation}
    \Omega_o = 0, \qquad \Omega_{\pm} = \pm i \abs{\alpha}\abs{k}.
\end{equation}
This implies the relations
\begin{eqnarray*}
    \Omega_o &=& - \vec{v}\cdot\vec{k} + i\beta\abs{k}^2 \nonumber \\
    \Omega_+ &=& - \vec{v}\cdot\vec{k} + i\abs{k}\left(\beta\abs{k} + \abs{\alpha}\right) \nonumber \\
    \Omega_- &=& - \vec{v}\cdot\vec{k} + i\abs{k}\left(\beta\abs{k} - \abs{\alpha}\right)
\end{eqnarray*}
As motivated in the introduction, well-posedness concerns the behaviour of the theory at high frequency. The above modes can be regarded as waves with different ``polarizations'', being the eigenvectors of the principal part essentially the polarization vectors of high frequency modes. Thus, in the high frequency limit ($|\vec{k}|\to\infty$) all roots have positive imaginary part, getting thus a well-posed behaviour of the equations. 

We finally notice that this result is also true even when: (i) $\alpha=\beta=0$; and (ii) $\alpha = 0$, $\beta \neq 0$. In both cases, the principal part turns out to be diagonal, with real eigenvalues, thus implying the system admits a well-posed initial value problem. In the next section we focus on a particular magnetic field scenario, and show an inequality involving $\alpha$ and $\beta$ which is necessary for the dynamo to work and make the magnetic energy grow in time. For this purpose, we apply our well-posedness results.

\section{Application: Energy bounds in Force-Free dynamos}
\label{sec-4}

We now apply the well-posed formulation discussed in section \ref{sec-3}, when the system is coupled with the so-called ``force-free'' condition. After an examination of the corresponding constraint propagation, we derive an identity from the study of the magnetic helicity, and give an estimate on the magnetic energy evolution. Finally, a simple inequality is shown as to be a necessary condition for an increasing of the magnetic energy as a consequence of the dynamo evolution equations.

\subsection{Dynamical equations}

Force-free fields have been deeply studied in past years \cite{Berger88,Sreenivasan73,Carrasco17,Rubio17}. They are useful for modeling strong magnetic fields surrounding compact objects like pulsars and black holes, or in regions where the electromagnetic field dominates over the plasma in an accretion mechanisms or gamma-ray bursts, resulting a decoupled dynamics. Even the solar corona can be locally described by force-free fields. Under this configuration, the electric field turns out to be everywhere orthogonal to the magnetic field, given that its parallel component vanishes due to free availability of charges. It can be also shown that the electric field is everywhere weaker than the magnetic field implying that, by invariance arguments, there always exists a local frame where fields are purely magnetic, and the current density flows along them.

The magnetic dynamo system of equations in the force-free regime reads
\begin{eqnarray} \label{ff-const}
    \partial_t\vec{B} &=& \vec{\nabla}\times (\vec{v} \times \vec B) + \vec{\nabla} \times (\alpha\vec{B}) + \beta\nabla^2\vec{B} \nonumber \\
    \vec{\nabla}\times\vec{B} &=& \gamma\vec{B}  \\
    \vec{\nabla}\cdot\vec{B} &=& 0 \nonumber
\end{eqnarray}
where $\gamma=\gamma(t,\vec{x})$. Notice that both constraint equations present in  the system (\ref{ff-const}) imply the new condition
\begin{equation}
    \vec{B}\cdot \vec{\nabla} \gamma = 0,
\end{equation}
that is, $\gamma$ must be constant along magnetic field lines during evolution. It is not a differential constraint, since it does not contain derivatives of $\vec{B}$. Nevertheless, it holds as a necessary condition for both differential constraints to satisfy during evolution, which in turn could imply that the magnetic field lines cannot intercept (unless $\gamma$ is a global constant or, at most, it is only a function of time).

\subsection{Constraint propagation}

As in the previous section, in which we analyzed how does the $C_1$ constraint propagate, the second equation in system (\ref{ff-const}) is known as the \textit{force-free constraint}, an extra condition whose propagation analysis shall be also taken into account. Here we prove that such a constraint does propagate correctly in time, as a consequence of the evolution equation of system (\ref{ff-const}).

Let us call the force-free constraint as
\begin{equation}\label{ff-constraint}
    \vec{C}_2 := \vec{\nabla}\times\vec{B} - \gamma\vec{B}.
\end{equation}
By differentiating with respect to time both sides of the above constraint and using the evolution equations for $\vec{B}$, one obtains an evolution equation candidate for $\vec{C}_2$. In order to prove that such equation has a unique solution for a given initial data, it is enough to analyze its principal part, which in this case reads

\begin{eqnarray}\label{evoC2}
    \partial_t\vec{C}_2 &=& \vec{\nabla}\times\vec{\nabla} \times(\vec{v}\times\vec{B}) - \gamma\vec{\nabla}\times(\vec{v}\times\vec{B}) \nonumber \\    
    &+& \alpha (\vec{\nabla}\times\vec{\nabla} \times \vec{B} - \gamma \vec{\nabla}\times\vec{B}) + \beta \nabla^2(\vec{\nabla}\times\vec{B} - \gamma\vec{B}). \nonumber \\
\end{eqnarray}
We now use the \textit{off-shell} identities
\begin{equation}
    \vec{\nabla}\times(\vec{v}\times\vec{C}_2) = \vec{\nabla}\times(\vec{v}\times(\vec{\nabla}\times\vec{B})) - \gamma\vec{\nabla}\times(\vec{v}\times\vec{B})
\end{equation}
and
\begin{equation}
    \vec{\nabla}\times\vec{C}_2 = \vec{\nabla}\times\vec{\nabla}\times\vec{B} - \gamma\vec{\nabla}\times\vec{B},    
\end{equation}
which just follow from the linearity property of vector cross product applied to equation (\ref{ff-constraint}). Then, equation (\ref{evoC2}) reduces to
\begin{equation}\label{final-const-C2}
    \partial_t\vec{C}_2 = \vec{\nabla}\times (\vec{v} \times \vec{C}_2) + \vec{\nabla} \times (\alpha\vec{C}_2) + \beta\nabla^2\vec{C}_2,
\end{equation}
which is exactly the same equation satisfied by the magnetic field. From our previous analysis of the corresponding initial-value problem, we conclude that equation (\ref{final-const-C2}) is well-posed and has therefore a unique solution for given smooth initial data. Thus, choosing $\vec{C}_2 = 0$ at $t = 0$, we conclude that $\vec{C}_2 \equiv 0$ for any further time, and the force-free constraint propagates correctly. 

\subsection{Energy estimates}
We now derive estimates on the magnetic energy in the force-free regime, for which we assume the function $\gamma$ to be locally constant and $\alpha > 0$. To that end, we first derive some useful results concerning the \textit{magnetic helicity}.

\subsubsection{Magnetic helicity}

Magnetic helicity quantifies various aspects of magnetic field structure \cite{Vishniac01}, being currently one crucial aspect for understanding astrophysical dynamos through numerical simulations \cite{Zhang18,Gosain19,Prabhu20,Brandenburg20Helicity}. It counts also for topological properties magnetic fields have as a consequence of the induction equation. It is a conserved quantity in Ideal MHD and approximately constant during magnetic reconnection. 

Starting from the Gauss linking number for two arbitrary smooths curves on $\mathbb{R}^3$ and by expressing the magnetic field as the curl of some vector potential 
\begin{equation} \label{eq-B-rotA}
\vec{B} = \vec{\nabla}\times\vec{A},   
\end{equation}
the magnetic helicity over a region $V\subseteq\mathbb{R}^3$ can be expressed as
\begin{equation}\label{helicity}
    \mathcal{H}_M = \int_{V}{\vec{A}\cdot\vec{B}}.
\end{equation}
Then, equation (\ref{ff-const}) implies that
\begin{equation}\label{eq-potA}
    \partial_t\vec{A} = \vec{v}\times\vec{B} + \alpha\vec{B} - \beta \vec{\nabla}\times\vec{B}.
\end{equation}
Taking a time derivative to expression (\ref{helicity}), we get
\begin{eqnarray*}
    \partial_t\mathcal{H}_M &=& \int_{V}{(\partial_t\vec{A})\cdot\vec{B}} + \int_{V}{\vec{A}\cdot(\partial_t\vec{B})}\\
    &=& \int_{V}{\alpha\,\vec{B}\cdot\vec{B}} - \beta\int_{V}{\vec{B}\cdot(\vec{\nabla}\times\vec{B})} + \int_{V}{\vec{A}\cdot(\vec{\nabla}\times\partial_t\vec{A})},
\end{eqnarray*}
where in the second line we used equations (\ref{eq-B-rotA}) and (\ref{eq-potA}). The third term of the right-hand side can be expressed as 
\begin{eqnarray*}
    \int_{V}{\vec{A}\cdot(\vec{\nabla}\times\partial_t\vec{A})} &=& \int_{V}{\varepsilon^{ijk}A_i\partial_j(\partial_t A)_k} \\
  &=& \int_{\partial V}{\varepsilon^{ijk} A_i n_j (\partial_t A)_k} - \int_{V}{\varepsilon^{ijk}\partial_j A_i(\partial_t A)_k} \\
    &=& \int_{\partial V}{\vec{A}\cdot(\hat{n}\times\partial_t\vec{A})} +  \int_{V}{\vec{B}\cdot(\partial_t\vec{A})},
\end{eqnarray*}
and it holds for any volume $V$. In particular, considering $V = S_R$ a ball of radius $R$, taking the limit $R\to\infty$ and using that $\vec{A}$ vanishes at infinity together with equation (\ref{eq-potA}), we arrive to the global identity
\[
    \int_{\mathbb{R}^3}{\vec{A}\cdot(\vec{\nabla}\times\partial_t\vec{A})} = \int_{\mathbb{R}^3}{\alpha\vec{B}\cdot\vec{B}} - \beta\int_{\mathbb{R}^3}{\vec{B}\cdot(\vec{\nabla}\times\vec{B})}.
\]
Thus, we finally obtain the relation
\begin{equation}\label{helicity-growth}
   \frac{1}{2}\partial_t\mathcal{H}_M = \int_{\mathbb{R}^3}{\alpha\,\vec{B}\cdot\vec{B}} - \beta\int_{\mathbb{R}^3}{\vec{B}\cdot(\vec{\nabla}\times\vec{B})}.
\end{equation}
This equality implies that, if $\alpha$ is a sufficiently large \textit{positive} function, the magnetic helicity would always increase. This property, nevertheless, does not necessarily tell us something about the global growth of the magnetic energy since, for instance, taking $\beta \ll 1$ (in appropriate units), we get $\partial_t\mathcal{H}_M \sim 2\alpha E_M$, from which we could have increasing magnetic helicity with constant magnetic energy. However, a relation between magnetic helicity and magnetic energy has been also noticed in the past \cite{DelSordo2010,Vishniac01,Berger99}.
Here we shall use identity (\ref{helicity-growth}) in order to give estimates for the magnetic energy, particularly in the force-free regime.

\subsubsection{Bounds on the magnetic energy and the mean helicity}

The force-free condition (\ref{ff-const}) implies that there exists a scalar function $f$ such that
\begin{equation}
    \vec{B} = \gamma \vec{A} + \vec{\nabla} f.
\end{equation}
Then, using the relation (\ref{helicity-growth}) and the constraints of system (\ref{ff-const}) we get
\begin{equation}
\int_{\mathbb{R}^3}{\alpha B^2} - \beta\gamma\int_{\mathbb{R}^3}{B^2} = \frac{1}{2\gamma}\partial_t\int_{\mathbb{R}^3}{\left(\vec{B} - \vec{\nabla} f\right)\cdot \vec{B}}
\end{equation}
The right hand side of the above equality can be expressed as
\begin{eqnarray*}
\frac{1}{2\gamma}\partial_t\int_{\mathbb{R}^3}{\left(\vec{B} - \vec{\nabla} f\right)\cdot \vec{B}}&=&
\lim_{R\to\infty}{\frac{1}{2\gamma} \left[\partial_t E_M(S_R) - \partial_t\int_{S_R}{\vec{\nabla}\cdot(f\vec{B})}\right]}\\
&=&  \lim_{R\to\infty}{\frac{1}{2\gamma}\left[\partial_t E_M(S_R) - \partial_t \int_{\partial S_R}{f\vec{B}\cdot\hat{n}}\right]}\\
&=& \frac{1}{2\gamma}\,\partial_t E_M,
\end{eqnarray*}
where in the last equality we have chosen $f$ to vanish at infinity. Now, using H\"older's inequality on $S_R$ we have
\begin{equation}
    \left|\int_{S_R}{\alpha B^2}\right|\leq \alpha_{\mbox{\scriptsize{max}}}(S_R)\int_{S_R}{B^2},
\end{equation}
where
\begin{equation}
    \alpha_{\mbox{\scriptsize{max}}}(S_R) = \max_{x\in S_R}{\abs{\alpha}}.
\end{equation}
Passing to the limit, we get
\begin{equation}\label{ine-final}
    \partial_t E_M \leq \left[2\gamma\left(\alpha_{\mbox{\scriptsize{max}}}- \beta\gamma\right)\right] E_M,
\end{equation}
with
\begin{equation}
    \alpha_{\mbox{\scriptsize{max}}} = \max_{x\in\mathbb{R}^3}{\abs{\alpha}},
\end{equation}
and using that $\alpha_{\mbox{\scriptsize{max}}}(S_R)\leq\alpha_{\mbox{\scriptsize{max}}}$, for any $R>0$. Inequality (\ref{ine-final}) can be integrated out in time, yielding
\begin{equation}\label{energy-bound}
E_M \leq E^{o}_M\,\exp{\left[2\gamma\left(\alpha_{\mbox{\scriptsize{max}}}- \beta\gamma\right)t\right]}    
\end{equation}
In particular, the magnetic energy may grow exponentially in time if and only if
\begin{equation}
    \alpha_{\mbox{\scriptsize{max}}} > \beta\gamma.
\end{equation}
Moreover, the equality in (\ref{energy-bound}) holds if and only if $\alpha$ is a positive constant.

\section{Final remarks}
\label{sec-5}

In this article, mathematical aspects of the system of equations describing the evolution of magnetic fields in a kinematic regime were addressed. 
In particular, we justified how it is possible to have growing modes (which are not purely ``physical'') without any dynamo-like mechanism. The underlying reason is the ill-posedness of the corresponding system of evolution equations. We illustrated this issue by providing two different configurations for the electromotive force: the first one being linear in the magnetic field, and the second one being linear in magnetic field derivatives. 

By studying the hyperbolicity of such formulations, we found that, in the first case, the theory is weakly-hyperbolic, implying that the system under this configuration does not constitute a well-posed initial-value problem. Moreover, there is no physical notion of energy for which the solution cannot be bounded in time with respect to the initial data. Thus, magnetic energy could reach arbitrarily large values, despite any dynamo-type mechanism. From the above results we conclude that this configuration should not be implemented or even considered, since growing linear perturbations may become arbitrary as the grid frequency is increased. Furthermore, non-linearities could alter such growth, making it to become exponential and spurious, thus leading to stiff numerical results. This kind of phenomena was already found in early days of dynamo theory. There have been cases of growing solutions of Ideal MHD equations that later turned out to be spurious numerical $\alpha$ terms by the lack of resolution, and where numerical solutions with no physical meaning have been noticed. Here we understood what the underlying mathematical problem is, suggesting the need of dissipative terms which can be re-interpreted by means of the hyperbolicity of the corresponding system of equations \cite{ReulaRubio17}. It is worthwhile to mention that the set of equations here addressed differ from the MHD equations, which actually have a well-posed initial-value formulation. For this reason, we consider that their ill-posedness is actually remarkable, given that they are often used to explain dynamo proceses in galactic and extragalactic scales \cite{Widrow02}. When coupling the induction equation to any set of well-posed fluid equations without any dependence on the magnetic field (as the Euler or Navier-Stokes equations), well-posedness only depends on the mathematical properties of the principal part of the “induction sector”. Moreover, as a consequence of Strang's theorem, we showed that if the dynamo equation with a fixed background velocity field is ill-posed, then coupling it with any set of fluid equations (without containing terms with derivatives of the magnetic field) will also give rise to an ill-posed system.

In the second case, instead, we proved that the theory is strongly-hyperbolic, implying that there exist a norm such that it is possible to bound the magnetic energy with the initial data. In this case, magnetic energy may increase exponentially in time, as a consequence of rather plausible dynamo-type mechanisms. We then applied this well-posed formulation to the force-free regime, which constitutes the configuration of minimal energy of magnetic fields. In particular, we studied the constraint propagation, and derived estimates for the magnetic energy, being able to prove an exponential growth in the case of constant mean helicity.

As a future perspective, it would be interesting to consider the full magnetic induction equation coupled with Hydrodynamics, taking into account contributions of the magnetic pressure in the fluid evolution equations (as in MHD), in order to see whether or not the problem of ill-posed is removed. Also, a study about turbulent effects and fluid-magnetic field fluctuations correlations when including further terms in the electromotive force (for instance, the cross-helicity term) would be worthwhile, as suggested in \cite{Marsch92,Marsch93}. In addition, we find it worth to understand the hyperbolicity problem of the same system of equations addressed in this work, as well as the corresponding extension to the relativistic regime, but from a different perspective, in particular considering the Leray-Ohya theory \cite{Leray53,Leray-Ohya67}. This approach focuses on the initial-value problem from initial data belonging to certain spaces of functions which admit topologies that do not come from any norm (e.g., the Gevrey classes; see \cite{Choquet-Bruhat52fr,Choquet08} for definitions, main results and applications). This alternative scenario seems to be rather natural for the study of viscous fluid equations, even in the full relativistic regime as shown, for instance, in \cite{Disconzi15,Bemfica17}.

As a general conclusion, a hyperbolicity analysis of the different theories carried out in order to describe magnetic field evolution and amplification mechanisms should be performed prior to make numerical simulations. This is a quite general consideration, being particularly relevant for the problem of the origins and evolution of cosmological magnetic fields.

\appendix

\section{Linear hyperbolic systems of equations}
\label{sec-ap}

In this appendix, we briefly review some basic concepts on linear and quasi-linear first-order systems of equations. In particular, we introduce some useful notions about hyperbolic first-order systems in a purely algebraic picture.

In order to get started, we first consider the simplest example, given by
\begin{equation}\label{eq:kreiss_systems-one}
{\left\{
{\begin{array}{rcl}
\partial_t u & = & A^i \partial_i u\; \\
u(0,x) & = & f(x)\;
\end{array}}
\right.}
\end{equation}
where $u:\mathbb{R}_{\geq 0}\times\mathbb{R}^3\to\mathbb{R}^{N}$ is a smooth vector field, $x = (x_1,x_2,x_3)$ are spatial coordinates, $\{A^i\}_{i=1}^3$ a set of real constant $N \times N$ matrices (being $N$ the number of dynamical fields encoded in $u$) and $f:\mathbb{R}^3 \to \mathbb{R}^N$ is a vector field. The \textit{Cauchy problem} or \textit{initial-value problem} for system (\ref{eq:kreiss_systems-one}) consists on finding a unique solution $u(t,x)$ satisfying a given initial data $u(0,x)=f(x)$. To this end, we give the following

\begin{definition}
\label{well-posed}
System (\ref{eq:kreiss_systems-one}) is called \textit{well-posed} if it admits a \textit{unique} solution in a neighborhood of $t=0$, and it \textit{continuously} depends on the initial data; that is, there exists a norm $\norm{\,\cdot\,}$ and a pair of real constants $C$, $\alpha$ such that, for all smooth initial data $f$ and any $t > 0$, the following inequality holds:
\begin{equation} \label{eq:desig-wellposed}
\norm{u(t,x)} \leq C e^{\alpha t} \norm{f(x)}.
\end{equation}
\end{definition}

The previous definition of well-posedness involves a subtle inequality which in general is not simple to verify. However, it is possible to characterize well-posedness by giving \textit{algebraic} conditions for the \textit{principal part} of the equation; that is, the part containing the derivatives of higher order. Particularly, some crucial results about well-posedness for constant-coefficient first order systems have been provided by Kreiss in \cite{Kreiss89}, thus reducing the the problem of well-posedness into a pure algebraic issue, as shown for instance in the following

\begin{theorem}\label{teo2}
System (\ref{eq:kreiss_systems-one}) is \textit{well-posed} in $L^2$ norm if and only if there exist constants $C$ and $\alpha$ such that, for all $t > 0$ and for all $k \in \mathbb{R}^n$,
\begin{equation}\label{ineq-sh}
\abs{e^{iA^jk_jt}} \leq C e^{\alpha t},
\end{equation} 
where $\abs{\,\cdot\,}$ is the usual matrix norm.
\end{theorem}

As an example, ideal Hydrodynamics constitutes a \textit{strictly-hyperbolic} system; that is, it is strongly-hyperbolic with all different real eigenvalues, corresponding to the propagation velocities of the fluid perturbations (see for instance \cite{Alcubierre08})\footnote{Although Hydrodynamics admits solutions which may develop a turbulent behaviour, one should not confuse such non-linear effects with the notion of well-posedness, in which only matters the \textit{principal part} of the system of equations.}.

Studying the \textit{hyperbolicity} of a theory means analyzing under which mathematical assumptions such conditions are verified. We also mention that the above definition is rather general, as there could exist quite peculiar norms satisfying inequality (\ref{eq:desig-wellposed}), which do not imply existence and uniqueness of solution in a stronger sense (i.e., with respect to certain other norms associated to functional spaces in which one generally expects a physical solution to belong). 

\subsection{Strong hyperbolicity}

There are several ways to introduce the concept of hyperbolicity. Intuitively, this idea is associated with some properties which are satisfied by systems that behave ``similarly'' to the wave equation, which has finite propagation speed of the information and thus, bounded (finite) domain of dependence. Although there are a few notions of hyperbolicity (some of them stronger than others), here we introduce the notion of \textit{strong hyperbolicity}.

\begin{definition}
\label{def1}

System (\ref{eq:kreiss_systems-one}) is called \textit{strongly-hyperbolic} if for any covector $k_i$, the matrix $A^i k_i$ is diagonalizable with only real eigenvalues.
\end{definition}

The symbol $A^i k_i$ is called the \textit{principal symbol} of system (\ref{eq:kreiss_systems-one}). From linear algebra, it is well-known that every complex matrix $A$ is diagonalizable with real eigenvalues if and only if there exists a \textit{symmetrizer} $H$, that is, a bi-linear and positive definite 2-form such that the composition $HA$ is \textit{symmetric}. Then, from Def. \ref{def1}, one can deduce that system (\ref{eq:kreiss_systems-one}) results strongly-hyperbolic \textit{if and only if} for each $k_i$ there exists a matrix $H(k)$ such that the composition $H(k)A$ is symmetric\footnote{Moreover, if such matrix $H$ does not depend on $k_i$; that is, if one can manage to use the same $H$ matrix for every $k_i$, the system is called \textit{symmetric-hyperbolic} \cite{Hormander97}. It is clear that symmetric hyperbolicity implies strong hyperbolicity, although the reciprocate is not necessarily true. For this subtle reason, since the main results shown in this work involve proving \textit{non-strong hyperbolicity} of the corresponding equations, no arguments against symmetric hyperbolicity were necessary to be included to this end.}.

Although the issue of finding a symmetrizer $H(k)$ (whenever it exists) generally results a non-trivial task, this is a useful criterion in order to check strong- hyperbolicity. Moreover, if $H$ is independent of $k$, it is possible to construct an inner product and thus a norm coming from it. In effect, by defining
\begin{equation}
\langle v, w\rangle := v^{\dagger} H w,
\end{equation}
we get $\norm{u}:=\sqrt{\langle u,u\rangle}$. Then, as a consequence of the symmetry of $HA$, the \textit{energy}
\begin{equation}
    E(t) = \int{\norm{u(t,x)}^2\,dx}
\end{equation}
is conserved during evolution. This simple calculation illustrates the relationship between well-posed systems and the possibility to associate a \textit{bounded energy} to them. 
 
\subsection{Ill-posedness}

If the system is such that the principal symbol $A^ik_i$ has real eigenvalues, but their eigenvectors \textit{do not} form a basis of $\mathbb{R}^3$ (that is, if $A^ik_i$ \textit{is not} diagonalizable), the system is said \textit{weakly-hyperbolic}, for which inequality (\ref{ineq-sh}) becomes weaker, namely
\begin{equation}\label{eq:w-hyp-ineq}
\abs{e^{iA^jk_jt}}\leq C\left[1 + (|\vec{k}|t)^{\beta}\right] e^{\alpha t},
\end{equation}
for real constants $C$, $\alpha$, $\beta\neq 0$ and $t\geq 0$.

These types of systems are characterized by having solutions that grow up to a polynomial in $|\vec{k}|t$, so they cannot be bounded independently of $|\vec{k}|$. Inequality (\ref{eq:w-hyp-ineq}) means that such a solution is a continuous function of the initial data but in different topologies (i.e., in Sobolev spaces of different orders). This does not turn out to be the desired situation if a numerical implementation is intended, since it would imply a loss of ``differentiability'' at each iteration, obtaining less and less smooth solutions. This problem can be traced down from the algebraic properties of the corresponding principal symbol, which in this case is a Jordan block of order 2, with two equal eigenvalues, even if not diagonalizable \cite{kreiss2004initial}. It is also common that the addition of perturbations to strongly hyperbolic systems with constant coefficients destroys the smoothness of the original solutions. A simple example of this case can be seen in \cite{kreiss2004initial} where the inclusion of lower-order perturbative terms causes an exponential growth of some frequencies of the solution in rather short times.

\subsubsection{Linear systems with variable coefficients}

All of the notions presented in the previous section can be successfully generalized to any linear first-order system with variable coefficients \cite{geroch1996partial}, namely
\begin{equation}\label{eq:kreiss_systems}
{\left\{
{\begin{array}{rcl}
\partial_t u^{\alpha} & = & P^{\alpha c}{}_{\beta}(t,x)\, \partial_c u^{\beta} + Q^{\alpha}(t,x)\; \\
u^{\alpha}(0,x) & = & f^{\alpha}\;
\end{array}}
\right.}
\end{equation}
where the dynamical fields $u^{\alpha} = u^{\alpha}(t,x)$ may be arbitrary tensor fields, and $P^{\alpha i}{}_{\beta}$ and $Q^{\alpha}$ are smooth functions of the coordinates\footnote{Furthermore, the notions of hyperbolicity and well-posedness can be generalized for \textit{quasi-linear} systems; i.e., systems of equations which are linear in their field derivatives, but admit coefficients that depend \textit{non-linearly} on the dynamical fields, as well as on products between these fields and their first derivatives. We shall not consider them in this context, as it goes beyond the scopes of this note.}. 
The most intuitive way to generalize our previous ideas in this case is by ``freezing out'' the function $P^{\alpha i}{}_{\beta}$ at some point, say $(t_o, x_o)$. By this way, it is possible to show that the notion of strong-hyperbolicity previously introduced implies that system (\ref{eq:kreiss_systems}) is locally well-posed in a neighborhood of $(t_o, x_o)$, using similar versions of Def. \ref{def1} and Theorem \ref{teo2} (see \cite{kreiss2004initial,geroch1996partial} for details). The main difference lies in the fact that existence and uniqueness results can only be reached \textit{locally} in time.

The magnetic induction system of equations fits in the set of linear equations with variable coefficients; that is, there always exist $P^{\alpha i}{}_{\beta}$ and $Q^{\alpha}$ such that the system can be put in the form (\ref{eq:kreiss_systems}) or, if it is a second (or higher)-order system, it can be always reduced to such a form by properly introducing extra fields. Nonetheless, its corresponding initial-value problem turns out to be non-trivial, depending on the choice of the electromotive force and deserves a careful analysis which, to the best of our knowledge it has not been addressed before.

\section*{Acknowledgments}

We thank A. Brandenburg, O. Reula and A. Esquivel for helpful conversations and insights. We also thank the anonymous referees for their comments and suggestions after a careful reading of the manuscript. This work was partially supported by grants PIP 11220130100365CO and PICT-2016-4174 from CONICET and FONCyT (Argentina); and by SECyT-UNC. M.E.R is a postdoctoral fellow of CONICET (Argentina).

\bibliographystyle{ieeetr} 
 \bibliography{dynamo}





\end{document}